\documentclass[final,numberedheadings]{aipproc}
\usepackage[latin1]{inputenc}
\usepackage[T1]{fontenc}
\usepackage{natbib}
\usepackage{amsfonts}
\usepackage{graphicx}
\usepackage{amsbsy}
\usepackage{amssymb}
\usepackage{amsmath}
\usepackage[usenames]{color}
\usepackage{bm}
\usepackage{latexsym}

\newcommand{\be}{\begin{equation}}
\newcommand{\ba}{\begin{eqnarray}}
\newcommand{\ea}{\end{eqnarray}}
\newcommand{\ee}{\end{equation}}

\layoutstyle{8x11single}

\begin{document}
\title{Feasibility Study of $\phi$(1020) Production at NICA/MPD}


\author{L.~S.~Yordanova and V.~I. ~Kolesnikov \\
for the MPD Collaboration}{
address={Veksler and Baldin Laboratory of High Energy Physics, \\
Joint Institute for Nuclear Research, \\
141980 Dubna, Moscow region, Russia \\ 
e-mail: \textcolor{blue}{\textup{kleo666@gmail.com}}}
}

\begin{abstract}
The goal of this article is to give information about the new accelerator complex NICA at JINR, to provide overview of the MultiPurpose Detector (MPD) and its subdetectors and to present the current results of the MPD performance for $\phi$-meson production. In our study we use the channel decay $\Phi\rightarrow  K^+ K^- $ to detect the formation of the $\phi$-meson. UrQMD event generator is used and central events at $\sqrt{s}$ = 11 GeV are analyzed.  The obtained peak from the invariant mass distribution is  fitted by a Breit-Wigner function. The calculated values of the parameters are consistent with the values given in literature. This study shows that the measurement of $\phi$-mesons is feasible at NICA/MPD.
\end{abstract}

\maketitle

\section{Introduction}

\begin{figure}
\centering 
\includegraphics[height=8cm]{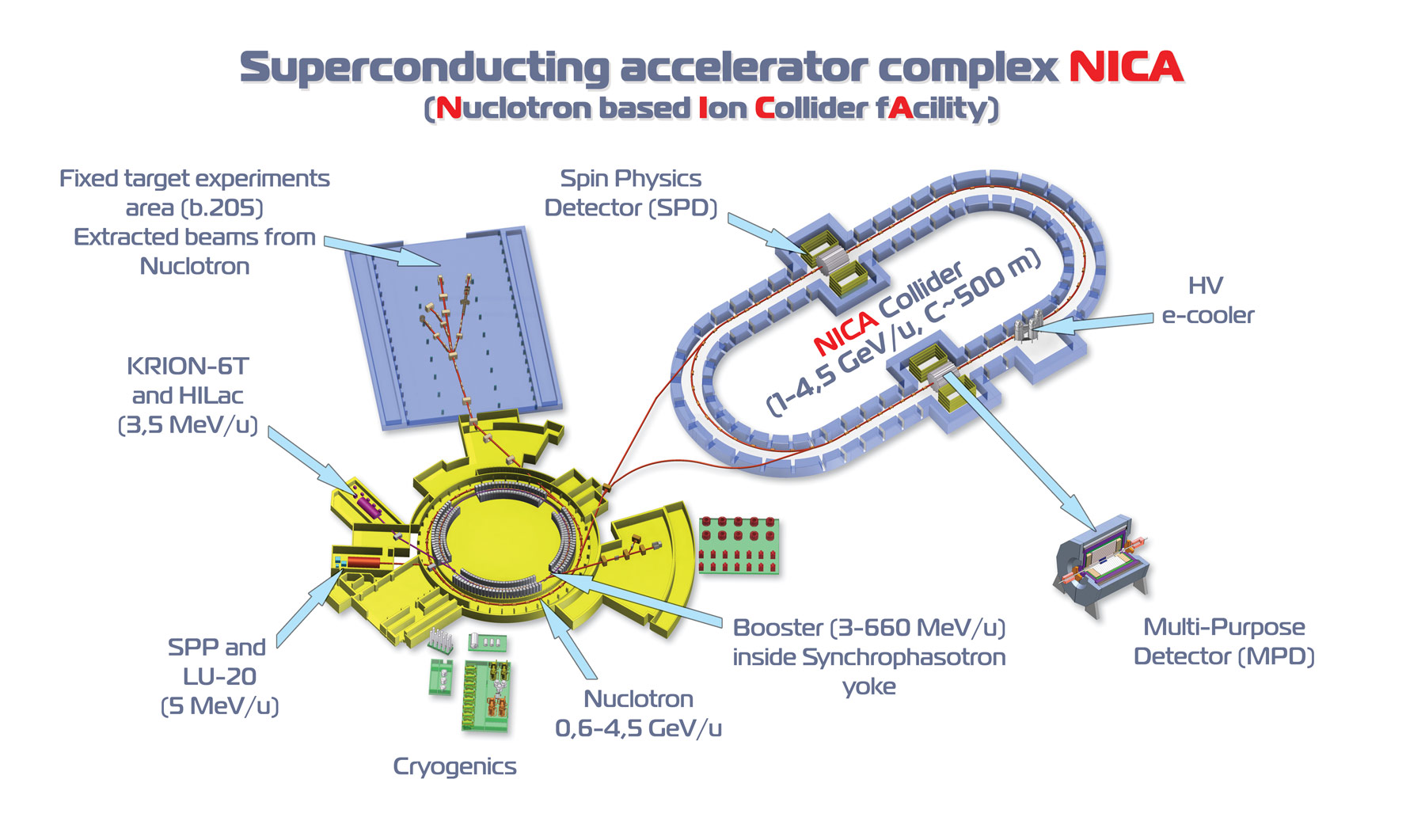}
\caption{NICA accelerator complex at JINR.}
\label{fig:fig1}
\end{figure}

\hskip 3mm The Nuclotron-based Ion Collider fAcility (NICA) is a new accelerator complex being constructed at JINR, Dubna, Russia (Fig.~\ref{fig:fig1}). The global scientific goal of the NICA/MPD project is to explore the phase diagram of strongly interacting matter in the region of highly compressed baryonic matter. The study of hot and dense baryonic matter provides relevant information on the in-medium properties of hadrons and nuclear matter equation of state; allows a search for deconfinement and/or chiral symmetry restoration, phase transition, mixed phase and critical end-point, possible strong P- and CP violation; gives information about the evolution of the Early Universe and the formation of neutron stars~\cite{NICA1}.

NICA's aim is to provide collisions of heavy ions over a wide range of atomic masses, from Au+Au collisions at $\sqrt{s_{NN}}$ = 11 GeV (for $Au^{79+}$) and an average luminosity of $L = 10^{27} cm^{-2} s^{-1}$ to proton-proton collisions with $\sqrt{s_{pp}}$ = 20 GeV and $L \sim 10^{32} cm^{-2} s^{-1}$. Two interaction points are foreseen at NICA which provide a possibility for two detectors to operate simultaneously - MultiPurpose Detector (MPD) and Spin Physics Detector (SPD). This overview is focused on the MPD detector. 

The MPD experimental program includes simultaneous measurements of observables that are presumably sensitive to high nuclear density effects and phase transitions.The goal to start energy scan as soon as the first beams are available and the present constraints in resources and manpower lead to the MPD staging. In the first stage of the project (starting in 2017) are considered - multiplicity and spectral characteristics of the identified hadrons including strange particles, multi-strange baryons and antibaryons; event-by-event fluctuations in multiplicity, charges and transverse momentum; collective flows (directed, elliptic and higher ones) for observed hadrons. In the second stage (starting in 2020) the electromagnetic probes (photons and dileptons) will be measured. It is proposed that along with heavy ions NICA will provide proton and light ion beams including the possibility to use polarized beams~\cite{NICA2}.
	
The software of the MPD project is responsible for the design, evaluation and calibration of the detector;  the storage, access, reconstruction and analysis of the data; and the support of a distributed computing infrastructure.The software framework for the MPD experiment (MpdRoot) is based on the object-orientated framework FairRoot (developed at GSI) and provides a powerful tool for detector performance studies, development of algorithms for reconstruction and physics analysis of the data~\cite{FAIR}. Extended set of event generators for heavy ion collisions is used (UrQMD, LAQGSM, HSD). 
	
\section{Overview of the MPD detector}

\begin{figure}
\centering 
\includegraphics[height=8cm]{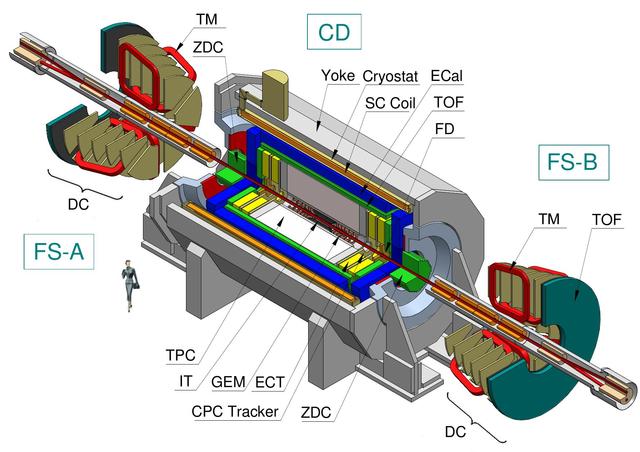}
\caption{General view of the MPD detector.}
\label{fig:fig2}
\end{figure}

\hskip 3mm The detector for exploring phase diagram of strongly interacting matter in a high track multiplicity environment has to cover a large phase space, be functional at high interaction rates and comprise high efficiency and excellent particle identification capabilities. The MPD detector matches all these requirements (Fig.~\ref{fig:fig2}). It consists of central detector (CD) and two optional forward spectrometers (FS-A and FS-B). The central detector consists of a barrel part and two end caps. The barrel part is a set of various subdetectors. The main tracker is the time projection chamber (TPC) supplemented by the inner tracker (IT). IT and TPC have to provide precise tracking, momentum determination and vertex reconstruction. The high performance time of flight (TOF) system must be able to identify charged hadrons and nuclear clusters in a broad pseudorapidity range. The electromagnetic calorimeter (ECAL) should identify electrons, photons and measure their energy with high precision. The zero degree calorimeter (ZDC) should provide event centrality and event plane determination, and also measurement of the energy deposited by spectators. There are also a straw-tube tracker (ECT) and a fast forward detector (FFD).

The magnet of MPD is a solenoid with a thin superconducting NbTi winding and a flux return iron yoke.  The magnet should provide a homogeneous magnetic field of 0.5 T.  The field inhomogeneity in the tracker area of the detector is about 0.1\%.

The MPD time projection chamber (TPC) is the main tracking detector of the central barrel and has to provide charged particles momentum measurement with sufficient resolution (about 2\% at pt=300 MeV/c), particle identification and vertex determination, two track separation (with a resolution <1 cm) and dE/dx measurement (dE/dx resolution better than 8\%) for hadronic and leptonic observables at pseudorapidities $|\eta|<2.0$ and pt >100 MeV/c.  TPC readout system is based on Multi-Wire Proportional Chambers (MWPC) with cathode readout pads.  
     
The inner tracker system (ITS) should enhance track reconstruction. ITS is able to restore tracks of particles with momentum less than 100 MeV/c and provides precise primary and secondary vertexes reconstruction with an accuracy of  $\sim 40 \mu m$. ITS should also identify relatively rare events with production of hyperons. ITS is based on the Silicon Strip Detector (SSD) technology. It consists of a silicon barrel and discs which will register particles with large pseudorapidity $|\eta| < 2.5$. 
     
The identification of charged hadrons (PID) at intermediate momentum (0.1-3 GeV/c) is achieved by the time-of-flight (TOF) measurements which are complemented by the energy loss (dE/dx) information from the TPC and IT detector systems. TOF system should provide a large phase space coverage $|\eta| < 3.0$, high combined geometrical and detection efficiency (better than 80\%), identification of pions and kaons with 0.1< pt < 2 GeV/c and  (anti)protons with 0.3 < pt < 3 GeV/c. The choice for the TOF system is Multigap Resistive Plate Counters (MRPC) which have good time resolution of $\delta < 70 ps$. The barrel covers the pseudorapidity region  $|\eta| < 1.5$ and the geometry efficiency in it is above 90\%. The end cap system covers the pseudorapidity region 1.5< $|\eta|$ <3.0.  
     
The fast forward detector (FFD) should provide fast determination of a nucleus-nucleus interaction in the center of MPD, generation of a start pulse for TOF and production of L0-trigger signal. The proposed FFD design is a granulated Cherenkov detector which has a high efficiency for the high energy photons and for ultra-relativistic charged particles as well. Its acceptance in pseudorapidity is 2.0$\leqq|\eta|\leqq$4.0. FFD has excellent time resolution of 38 ps. 
     
The end cap tracker (ECT) has to provide charged particle identification and momentum measurement in the pseudorapidities 1 < $|\eta|$ < 2.2. ECT has high track reconstruction efficiency and momentum resolution about 10\%. ECT consists of two end cap parts which are made with modules, containing layers of straw tubes. 
     
The primary role of the electromagnetic calorimeter (ECAL) is to measure the spatial position and energy of electrons and photons. ECAL should have a high segmentation, should provide good space resolution, energy resolution about $3\%$ and should allow a separation of overlapping showers. The Pb-scintillator ECAL of the "shashlyk" type will be used. 
     
The zero degree calorimeter (ZDC) should provide a classification of events by centrality, event plane determination and measurement of the energy deposited by spectators. ZDC consists of modules of 60 lead-scintillator tile "sandwiches" with lead and scintillator plates. The light readout is provided by the wave-length shifting fibers (WLS-fibers) and the micropixel avalanche photodiodes (MAPDs).
 
More detailed description of the MPD detector can be found in the 'MPD Conceptual Design Report'~\cite{MPD}.

\section{Motivation for study of the $\Phi$-meson}

\hskip 3mm The $\phi$ vector meson is the lightest bound state of hidden strangeness, consisting of a quark-antiquark pair. Although it is a meson, it is heavy in comparison with mesons consisting of u and d quarks, having a mass ($m_{\phi} = 1019.456 \pm 0.020 MeV/c^2$) comparable to the proton and $\Lambda$ baryons. The $\phi$-meson is expected to have a very small cross-section for interactions with non-strange hadrons. Its observables should remain largely undisturbed by the hadronic rescattering phase  of the system's evolution. The $\phi$-meson  has a relatively long life-time of $\sim$ 46  fm/c which means that it will mostly decay outside the fireball and its decay daughters won't have enough time to rescatter in the hadronic phase. Previous experimental measurements of the  $\phi$/K- ratio as a function of centrality showed the possibility of  $\phi$ production via $K^+ + K^-$ coalescence in the hadronic stage. The observed ratios are flat as a function of centrality. These properties make the  $\phi$-meson an excellent probe of the hot and dense medium created in nucleus-nucleus collisions~\cite{Sarah, MEA, PHENIX}.

	Measurements of the production of strange particles such as the $\phi$-meson can provide important information on the properties of the medium and particle production mechanisms in ultra-relativistic Au-Au collisions. Measurements of the $\phi$-meson pT spectra and their dependence in terms of shape and normalization on centrality may shed light on the constituents of the medium at the time of $\phi$ formation as well as the mechanism through which the $\phi$-mesons are formed. Further insight into mechanisms of particle production for strange particles compared to non-strange particles can be gained through measurement of the particle ratios of multistrange hadrons. The medium produced at NICA/MPD can be also probed by measuring the elliptic flow of the $\phi$-meson. Since multistrange hadrons and particles with hidden strangeness are assumed to freeze out early and undergo fewer interactions in the hadronic stage, their v2 signals should provide a clean signal from the early stage of the system's evolution.  Measurements of the $\phi$-meson  v2  can give information on the collectivity and possible deconfinement of the system in the early stage and can serve to constrain different dynamical models of elliptic flow and particle production~\cite{SHOR, STRANGE}.
	
	The $\phi$-meson with its mass comparable to  $\Lambda$ and p and low interaction cross-section can be exploited as a very good tool to probe the properties of the medium produced in nucleus-nucleus collisions at NICA/MPD.  $\Phi$-meson observables may help to distinguish further between different physical models of the dense medium and to enrich our knowledge of the system created in ultra-relativistic heavy-ion collisions.

\section{Data set}

\hskip 3mm In our study we use the channel decay $\Phi\rightarrow  K^+ K^- $ to detect the formation of the $\phi$-meson. This channel is chosen because it has a high branching ratio (49.1\%) and kaons are easy to detect. UrQMD event generator is used and gold-gold collisions are generated~\cite{URQMD}. The number of the analyzed central events is 50 000 at $\sqrt{s}$ = 11 GeV. The detector configuration used  in the study includes the Time Projection Chamber (TPC) and the Time-Of-Flight system (TOF)  for track reconstruction and particle identification in the pseudorapidity range $|\eta|$ < 1.5.

\section{Reconstruction of $\Phi$-mesons }
\subsection{Signal distribution}

The study of  $\phi$-meson production in central  Au-Au collisions at $\sqrt{s}$ = 11 GeV is performed using the UrQMD event generator.  After the generation the events samples are transported through the MPD detector using the Geant 3 and  Geant 4 transport packages.

In this analysis, $\phi$-mesons are measured through the channel decay $\Phi\rightarrow  K^+ K^- $  . The kaon  daughter particles are identified through their ionisation energy loss (dE/dx ) in the Time Projection Chamber (TPC) and also by  using the information from the Time-Of-Flight system (TOF). A selection of kaon pairs by track quality cuts and particle identification (PID) is performed. The invariant mass of the kaon pairs is calculated by combination of all $K^+$ and $K^-$ pairs of the same event. Consequently, we obtain data which contain the signal of $\phi$-meson with the background. It is called same-event or signal distribution.

\subsection{Background estimation}

Since not all charged kaons in each event originate from $\phi$-meson decays, the $\phi$-meson signal extracted by this way is above a large combinatorial background of uncorrelated pairs. This combinatorial  background should be removed from the signal distribution. There are two methods for removing background events in case of mass reconstruction: mixed-event technique and same-event technique~\cite{Rand, MEA1}. In mixed-event technique to separate the signal from the background, pairs of $K^+$ and $K^-$ are generated from different events. Such pairs create the background and pairs of $K^+$ and $K^-$ from the same event give the signal. In same-event technique the background is generated from the pairs of $K^+ K^+$  and  $K^- K^-$  of the same event, while $K^+ K^-$  pairs of the same event give the signal. In our analysis we use the mixed-event technique and the combinatorial background is estimated from the uncorrelated  $K^+ K^-$  pairs. An invariant mass distribution is constructed using all positively charged kaon candidates from one event mixed with all negatively charged kaon candidates from $n$ other events where $n$ can be chosen relatively high in order to reduce effects from statistical fluctuations.

\subsection{Extraction and scaling of the combinatorial background}

The mixed-event background distribution is estimated by using a big number of events and it should be scaled before being subtracted from the signal distribution. The scaling can be performed by two different ways -  scaling by the integral ratio and scaling iteratively. ~\cite{Sarah} In the scaling by the integral ration, the background distribution is scaled by the ratio of the integrals of the signal and the background distributions in a defined mass region. The mass region should not include the $\phi$-meson mass peak. In the iterative scaling,  the background distribution is scaled by the ratio of the integrals of the signal and the background distributions in a fixed mass region including the $\phi$-meson mass peak. The background distribution is then subtracted from the signal distribution  and the remaining signal is fitted with Breit-Wigner function plus a straight line. The signal integral is the integral in the mass range minus the integral of the  Breit-Wigner function. Then the ratio is recalculated and the background is rescaled. After scaling the background distribution, the raw $\phi$-meson yields are obtained by subtracting the scaled mixed-event background distribution from the signal distribution.

\subsection{Breit-Wigner function}

The obtained peak from the invariant mass distribution after the subtraction is then fitted by a Breit-Wigner function and the characteristics of the $\phi$-meson such as its mass and its width are found. The Breit-Wigner distribution is a continuous probability distribution. It is most often used to model resonances (unstable particles)in high energy physics. The  Breit-Wigner function has the following appearance:

\begin{equation}
\quad BW_{(m_{inv})} = \frac{1}{2\pi} \frac{AW}{(m-m_\phi)^2 +(\frac{W}{2})^2} ,
\end{equation}

In this function A is the area of the distribution and W is the width. The measured values of the $\phi$ -meson mass and width are consistent, within the experimental resolution, with the particle data grop (PDG) values. The recent results of the $\phi$-meson study are shown in the next section. This study shows that the measurement of $\phi$-mesons is feasible and we can expect detection of them when NICA/MPD will be put in operation.

\subsection{Results}

\begin{figure}
\centering 
\includegraphics[height=6cm]{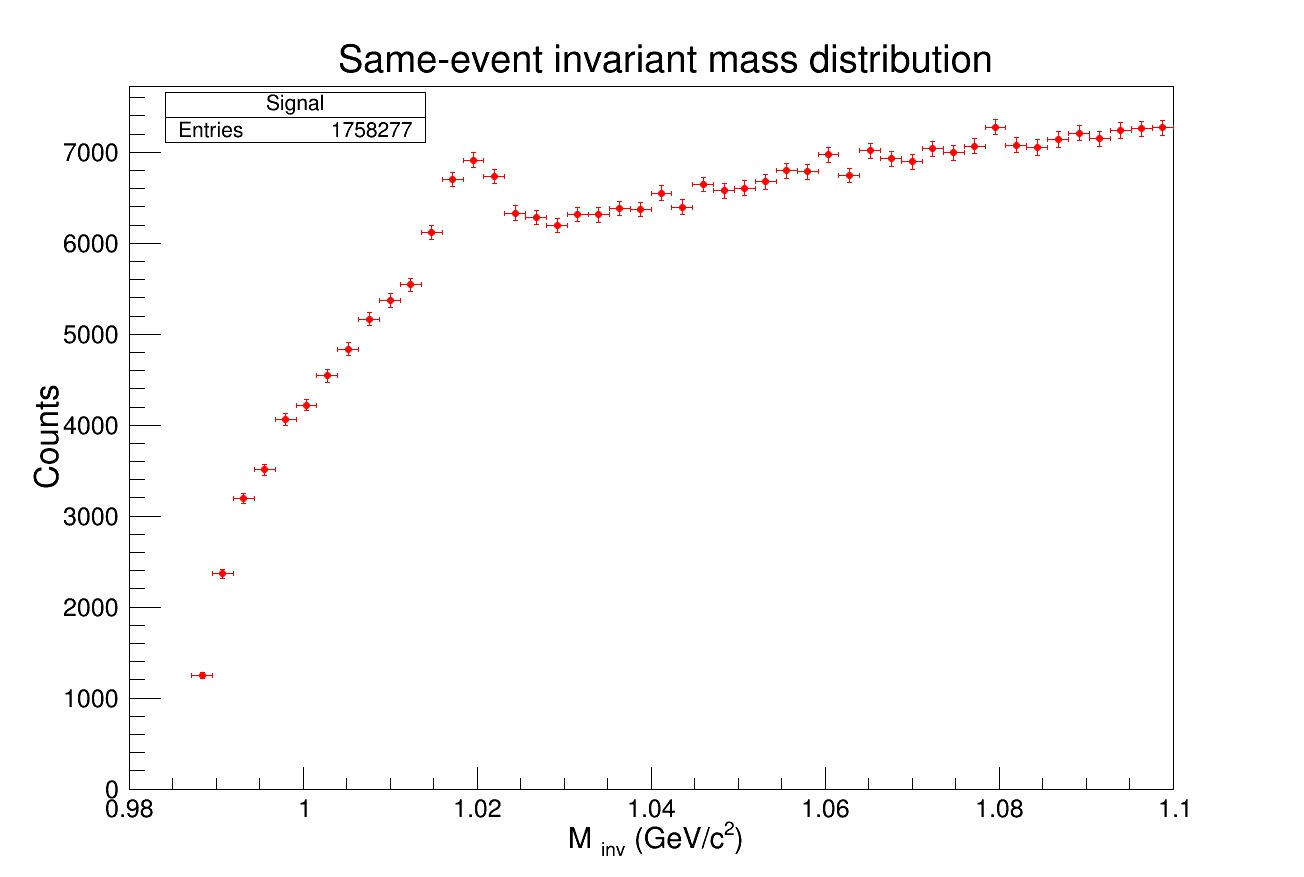}
\caption{Signal distribution.}
\label{fig:fig3}
\end{figure}

\begin{figure}
\centering 
\includegraphics[height=6cm]{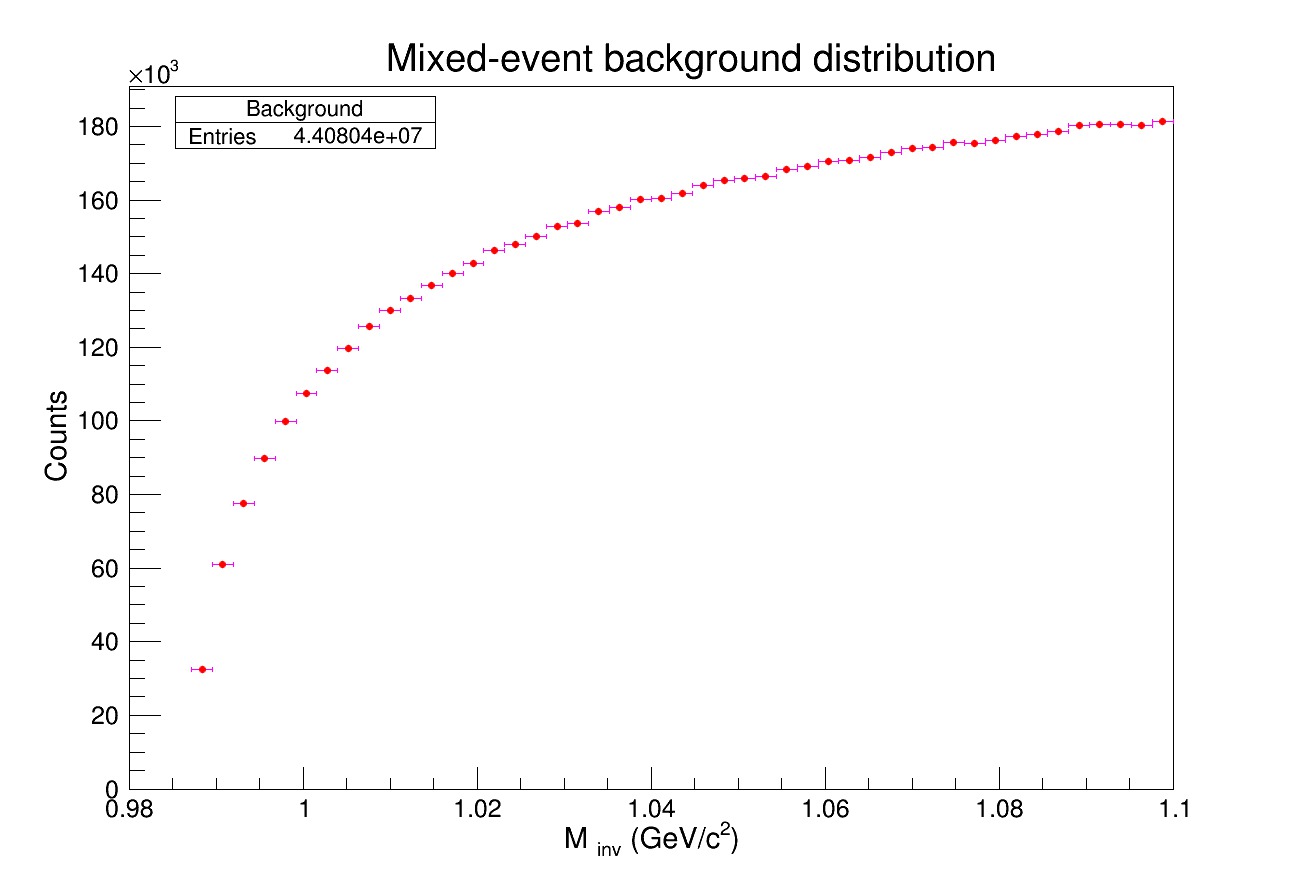}
\caption{Mixed-event background distribution.}
\label{fig:fig4}
\end{figure}

\begin{figure}
\centering 
\includegraphics[height=6cm]{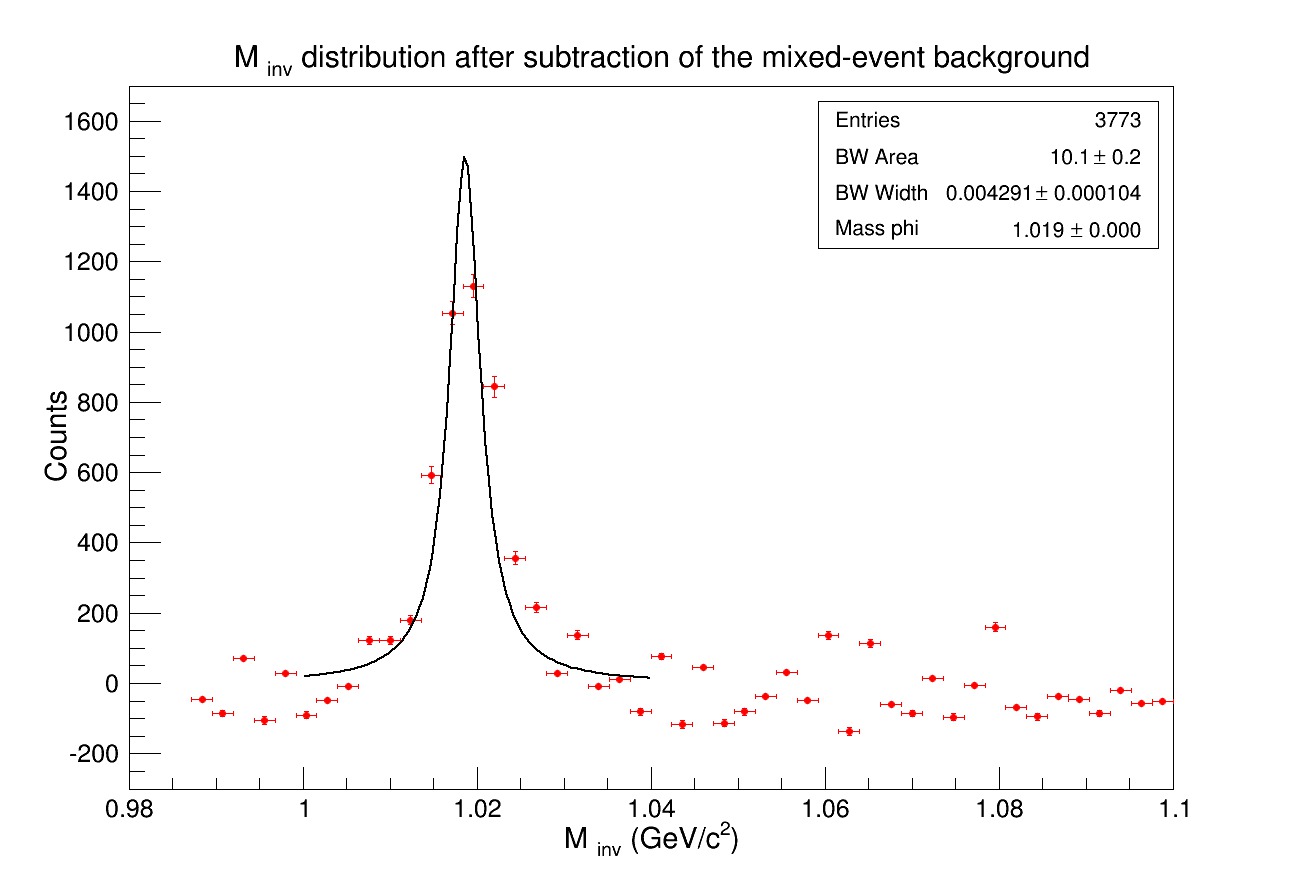}
\caption{$\Phi$-meson invariant mass distribution after subtraction of the mixed-event background.}
\label{fig:fig5}
\end{figure}

The recent results of the $\phi$-meson study are shown in the following histograms (see Fig.~\ref{fig:fig3}, Fig.~\ref{fig:fig4} and Fig.~\ref{fig:fig5}). 

In Fig.~\ref{fig:fig3} an invariant mass signal distribution of the $\phi$-meson is represented.  The exact values of selection cuts for the kaons used for the reconstruction of the $\phi$-mesons are found by performing a scan over the whole set of selection criteria with a requirement to maximize the invariant mass peak significance. The invariant mass peak significance is defined as $S/\sqrt{(S+B)}$ where $S$ and $B$ are total numbers of signal and background combinations inside $\pm 3\sigma$ interval around the peak position. For the current signal distribution it is - $S/\sqrt{(S+B)} = 18.11$. It is still seen that the peak of the signal is above a large combinatorial background, due to the uncorrelated kaon pairs.

Fig.~\ref{fig:fig4} shows a background estimation by using the mixed-event technique. The values of selection cuts correspond to those for the signal distribution. The background distribution has a smooth shape because of the relatively high number of events $n$ chosen for calculating it.

The final results for the raw yields of the $\phi$-meson are shown in Fig.~\ref{fig:fig5}. The values of the parameters obtained by the Breit-Wigner fit (Width = 4.291 $\pm$ 0.104 $MeV/c^2$ and $M_{inv}$ = 1019.995 $\pm$ 0.022 $MeV/c^2)$ are consistent with the values given in literature. All these preliminary results show that the measurement of $\phi$-mesons is feasible at NICA/MPD. The next step of the $\phi$-meson study is connected with detailed analyses of higher statistics and  measurements of the elliptic flow. The properties of the $\phi$-meson will be also studied in different types of collisions with different centrality.

\newpage
\section{Summary}

\hskip 3mm In conclusion, it should be said that the MPD detector has many advantages and meets all the ambitious physics requirements for exploring phase diagram of strongly interacting matter in a high track multiplicity environment. The MPD detector's advantages comprise coverage of a large phase space, functionality at high interaction rates, high efficiency and excellent particle identification capabilities. 

Measurements of the production of strange particles such as the $\phi$-meson can provide important information on the properties of the medium and particle production mechanisms in ultra-relativistic collisions. The $\phi$-meson with its mass comparable to  $\Lambda$ and p and low interaction cross-section can be used to probe the properties of the medium created at NICA/MPD. The study of the $\phi$-meson, based on the current analyses, is feasible and gives quite good results. More detailed study of the $\phi$-meson properties is included in the MPD physics programme. Therefore, NICA facilities provide unique capabilities for studying fundamental properties of the theory of strong interactions (QCD).

\bibliographystyle{aipproc}

\end{document}